\documentclass[preprints,article,accept,moreauthors,10pt,a4paper]{mdpi} 

\firstpage{1} 
\articlenumber{x}
\doinum{10.3390/------}
\pubvolume{xx}
\pubyear{2017}
\copyrightyear{2017}
\history{Received: date; Accepted: date; Published: date}

\pdfoutput=1

\usepackage{calrsfs}
\usepackage{amsfonts}
\usepackage[abs]{overpic}
\Title{Light-like shockwaves in scalar-tensor theories}

\Author{Bence Racsk\'{o}$^{1,}$* and L\'{a}szl\'{o} \'{A}rp\'{a}d Gergely$^{2}$}

\AuthorNames{Bence Racsk\'{o} and L\'{a}szl\'{o} \'{A}rp\'{a}d Gergely}

\address[1]{%
$^{1}$ \quad Institute of Physics, University of Szeged, Hungary; daeron806@gmail.com
\\
$^{2}$ \quad Institute of Physics, University of Szeged, Hungary; laszlo.a.gergely@gmail.com}

\corres{Correspondence: daeron806@gmail.com; Tel.: +36305535764}

\abstract{
Both electromagnetic shock-waves and gravitational waves propagate with the
speed of light. If they carry significant energy-momentum, this will change
the properties of the space-time they propagate through. This can be
described in terms of the junction conditions between space-time regions
separated by a singular, null hypersurface. We derived generic junction
conditions for Brans-Dicke theory in the Jordan frame, exploring a formalism
based on a transverse vector, rather than normal, which can be applied to
any type of hypersurfaces. In the particular case of a non-null hypersurface
we obtain a generalised Lanczos equation, in which the jump of the extrinsic
curvature is sourced by both the distributional energy-momentum tensor and
by the jump in the transverse derivative of the scalar. In the case of null
hypersurfaces, the distributional source is decomposed into surface density,
current and pressure. The latter however ought to vanish by virtue of the
scalar junction condition.}

\keyword{space-time junction conditions; Brans-Dicke theory; null hypersurfaces}

\begin{document}

\section{Introduction}

General relativity (GR) has withstood the confrontation with observations
both in Solar System tests and in strong field regimes, last of which had
been the experimental detection of gravitational waves by the LIGO
Scientific Collaboration and Virgo Collaboration {\cite{GW150915,GW151226,GW170104,GW170608,GW170814,GW170817}}.
The direct detection of gravitational waves from black hole binaries and a
neutron star coalescence has confirmed (through the test of the dispersion
relations and of the model dependent delay in the arrival time of the gamma radiation following the neutron star merger) that they propagate with the speed of light {\cite%
{GW170104,GW170817,GRtests,GWGRB}}. In consequence many modified/extended
theories of gravitation were disruled. Nevertheless there are still many
interesting such theories, which allow for light-like gravitational wave
propagation, still worth investigating.

General relativity (GR) has given predictions on both galactic scales and
beyond which can be reconciled with observations only at the price of
introducing still undetected (otherwise than gravitationally) dark matter
and dark energy. There is hope that modified gravity theories might replace
them by corresponding geometrical effects. Such modified gravity theories
encompass either a more complicated (higher-order) dynamics for the metric
tensor (but this may lead to instabilities and ghosts), or increase the
number of the fields describing pure gravity, with adding scalars, vectors,
2-form fields or even a second metric. The main difference as compared to
models with additional fields representing the dark sector is in the way the
metric couples to them: for the dark sector the coupling is minimal, for a
geometric field it may be more complicated. Simplest of them would be a
scalar-tensor theory. Horndeski has established \cite{Horndeski,DGSZ} the
most generic class of such theories with both the metric tensor and the
scalar field obeying second-order dynamics.

Historical interest in scalar-tensor theories of gravity began with the
Kaluza-Klein theory. In 1919 Kaluza sought to unify gravity with
electrodynamics by considering a five-dimensional spacetime, whose metric is
subject to the Einstein field equations. To account for the observed
four-dimensional nature of space-time, he assumed that the extra dimension
is compact and small, hence the dependence on this fifth coordinate could be
averaged out. Then, the five dimensional metric is decomposed into a four
dimensional metric (describing gravity), a four-vector (describing
electromagnetism), and a scalar field. At the time, the scalar field was
thought to be undesirable, however later on (especially when the connection
between fields and particles had been established), Kaluza-Klein theory
proved to be an inspiration for more general theories of gravity with added
scalar fields, such as the much-investigated Brans-Dicke theory.

In scalar-tensor theories it is customary to explore one of two conformally
related metrics $g_{\mu \nu }^{\prime }\left( x\right) =\Omega ^{2}\left(
x\right) g_{\mu \nu }\left( x\right) ~$(with $\Omega \left( x\right) $ a
nowhere-vanishing smooth function): 1) the Jordan (Jordan-Fierz or string) 
frame $g_{\mu \nu }$, in
which the scalar field $\phi$ is coupled non-minimally to gravity, but is not
coupled to matter fields, and 2) the Einstein frame $g_{\mu \nu }^{\prime }$%
, in which the coupling between gravity and the scalar field is minimal, but
there is an anomalous coupling of the scalar field to the matter fields \cite{CapozzielloFaraoni}.

Despite scalar-tensor theories being around for a
long time, there is still a heated debate on whether both frames are
physical or not, and if yes, whether they are physically equivalent. Dicke
argued \cite{Dicke} that since physics must be invariant under the rescaling
of units, and the conformal transformation is merely a local rescaling of
distances, therefore, physics should not depend on the conformal frame,
provided that the units of length, time and mass scale appropriately \cite{FaraoniNadeau}, although this argument has been criticized \cite{CapozzielloFaraoni}. Some authors argue in favour of the Einstein frame,
as in it the energy conditions for the scalar field are obeyed, while in the
Jordan frame the scalar field violates all known energy conditions \cite{FaraoniGunzig,MagnanoSokolowki}, also comparison with pure GR 
results is easier \cite{BarrabesBressange}. Other authors prefer the Jordan
frame, satisfying the equivalence principle, which is violated in the
Einstein frame due to the anomalous scalar field-matter couping. Therefore
in the Einstein frame the matter stress-energy tensor rather than obeying a
continuity equation is subject to $\nabla _{\mu }T^{\mu \nu }=-T\nabla ^{\nu
}\ln \Omega $ \cite{BarrabesBressange,CapozzielloFaraoni}.

Both in GR and in modified gravity theories it is of special interest to
match spacetime regions with different matter sources or even different set
of symmetries. This can be done along a common hypersurface, which may be
temporal, spatial or even null. Moreover, the hypersurface may contain a
distributional energy-momentum layer, complicating the junction conditions.
For GR they were worked out covariantly by Israel \cite{Israel}, but this
formalism does not apply for null hypersurfaces. In order to deal with them,
Barrab\`{e}s and Israel proposed a modified junction formalism \cite{BI},
relying on the use of a transverse vector to the null hypersurface.

Junction conditions can also be derived by employing a variational
principle, both for GR and scalar-tensor theories in the Einstein frame (incorporating the scalars in the matter sector) \cite%
{Mukohyama}. The dynamics of bubbles (infinitesimally thin shells) or plane domain walls were considered in Brans-Dicke theory in the Jordan frame \cite{SchmidtWang,LetelierWang,SakaiMaeda1,SakaiMaeda2}. For the Horndeski class of theories they were discussed in the Jordan frame in \cite{PS}, nevertheless only for space-like or time-like hypersurfaces and
further applied in a cosmological setup \cite{NishiKobayashi}. These results
however cannot be applied for null hypersurfaces. Such hypersurfaces may be
of physical interest as they represent light-like shock-waves, both
electromagnetic or gravitational, which modify the gravitational properties
of spacetime they propagate through.

In this paper, we investigate a general approach for the junction conditions
for scalar-tensor theories, which can be applied for any type of
hypersurfaces. We opt for the Jordan frame, motivated by the desire to keep the generic 
form of the function $G_{4}$ in the Hordeski Lagrangian. Another
motivation for assuming that the sources couple to gravity only via the
metric tensor, and not via the scalar field would be to avoid any
non-gravitational interaction of the scalar field with the baryonic matter
fields $\psi^{i}$, hence to be able to describe the dark sector with $\phi$. We
derive the Euler-Lagrange equations for both
the metric and the scalar and investigate the singular
contributions, which would appear only in the second derivatives of both
fields (terms proportional to Dirac-delta functions). This is related to the
approach of \cite{GerochTraschen}, which considers metrics whose
curvature tensors are well defined as distributions (e.g. avoid products of
distributions).

We derive the junction equations by separating the singular contributions to
the field equations and connecting them to distributional sources in Section
2. As a first exercise we apply the formalism for the Brans-Dicke
scalar-tensor theory in Section 3. We note that a related treatment was
developed in \cite{BarrabesBressange}, but in Einstein frame and for
multi-scalar fields. Finally in Section 4 we specify our results for null
hypersurfaces, which is followed by a Summary.

\section{Junction conditions across arbitrary hypersurfaces}

We consider the spacetime $M$ cut into two disjoint parts $M^{+}$ and $%
M^{-}$ by a common boundary hypersurface $\Sigma$, with unspecified causal
character. We allow certain otherwise smooth geometric quantities to undergo
sudden changes at the boundary, leading to discontinuities across $\Sigma$.
Physically, $\Sigma$ may separate a star from its exterior (collapsing stars
included), however it also can be the world-volume of a shockwave (for
example, one emanating from a supernova explosion), or a space-like
hypersurface encompassing a cosmological phase transition, among others.

Despite the junction, $M$ does possess a smooth structure \cite{ClarkeDray},
hence in principle it is possible to use a coordinate system that
transitions smoothly across $\Sigma$. However in practical situations, such
coordinate systems might not be straightforward to identify, hence we will
express all equations on the junction surface in coordinate charts internal
to $\Sigma$. In other words we use a doubly covariant formalism.

Null hypersurfaces representing shockwaves travelling at the speed
of light are physically relevant, nevertheless their study is obstructed by the fact that they have degenerate metrics and their
normal vectors are also tangential, preventing a proper orthogonal
decomposition of quantities along $\Sigma$. Therefore following \cite{BI} we
explore an oblique decomposition, valid for all types of hypersurfaces. The
holonomic basis vectors of the hypersurface are denoted $e_{a}^{\mu}$, with
a transverse vector field $l^{\mu}$ completing the basis. The normal
covector field $n_{\mu}$ satisfies $n_{\mu}l^{\mu}=1/\eta$ ($\eta$ arbitrary
and nonvanishing) with the norm $\epsilon=n^{\mu}n_{\mu}$ (depending on the
type of hypersurface, $\epsilon=\pm1,0$). If $\Sigma$ is given as the zero
set of a scalar field $f$, then $n_{\mu}=\frac{1}{\alpha}\partial_{\mu}f$,
with $\alpha$ a normalising factor.

For an arbitrary field quantity $F$ on $M$, its jump, arithmetic mean and
soldering accross $\Sigma$ are {\cite{BI}}: 
\begin{align}
\left[F\right] & =\left.F^{+}-F^{-}\right\vert _{\Sigma}~,  \notag \\
\bar{F} & =\frac{1}{2}\left.\left(F^{+}+F^{-}\right)\right\vert _{\Sigma}~, 
\notag \\
\widetilde{F} & =F^{+}\Theta\left(f\right)+F^{-}\Theta\left(-f\right)~,
\end{align}
with 
\begin{equation}
\Theta\left(x\right)=\left\{ 
\begin{array}{c}
1,\ x>0 \\ 
0,\ x<0 \\ 
\frac{1}{2},\ x=0%
\end{array}%
\right.
\end{equation}
the Heaviside function. It is straightforward to derive the relation 
\begin{equation}
\partial_{\mu}\widetilde{F}=\widetilde{\partial_{\mu}F}+\left[F\right]%
n_{\mu}\alpha\delta\left(f\right)~.
\end{equation}

In a scalar-tensor theory with at most second-order dynamics (Horndeski
class, \cite{Horndeski}, avoiding Ostrogradsky-instabilities {\cite%
{Ostrogradsky}, see also \cite{Woodard}}) we require that none of the
contributions to the field equations exhibit derivatives of Dirac-delta
functions (difficult to interpret from a physical point of view), while the
Dirac-delta functions themselves will be allowed (related to the density of
some finite quantity characterising an idealised, infinitely thin layer
along $\Sigma$). This condition can be assured by assuming both $g_{\mu\nu}$
(in smooth coordinates) and $\phi$ continuous across $\Sigma$. The
continuity of the induced metric is the first junction condition of Israel 
\cite{Israel}. The continuity of the rest of the metric components can be
assured by{\ picking up }$C^{1}${\ Gaussian normal coordinates, the
existence of which has been proven in \cite{ClarkeDray}}.

The action of the system is 
\begin{eqnarray}
S\left[g_{\mu\nu},\phi,\psi^{i}\right] & = & S_{G}\left[g_{\mu\nu},\phi%
\right]+S_{M}\left[g_{\mu\nu},\psi^{i}\right]~,  \notag \\
S_{G}\left[g_{\mu\nu},\phi\right] & = & \int
d^{4}x\left(\sum_{k=2}^{5}L_{k}\right)~,
\end{eqnarray}
with the gravitational part given by the Horndeski Lagrangians and the
matter part independent of $\phi$ in order to assure that the equivalence
principle remains valid (this argument applies to the Jordan frame, which
is the physical frame, where energy-momentum conservation holds).

We note that the recent confirmation of the gravitational wave propagation speed to agree with the speed of light at the order of one part in quadrillionth \cite{GRtests,GWGRB} at low redshifts has disruled theories with dependence of the kinetic term $X=-\frac{1}{2}\nabla^\mu\phi\nabla_\mu\phi$ in the coupling of the Ricci curvature $R$ and Einstein tensor $G_{\mu\nu}$ in $L_4$ and $L_5$, respectively \cite{DeFeliceTsujikawa,KobayashiYamaguchi}. Further, the latter does not depend on $\phi$ either (except through its derivatives), hence, due to the Bianchi identities, the whole $L_5$ ought to vanish \cite{Many} (see also \cite{EZ,CreminelliVernizzi}).

The energy-momentum tensor associated with the matter fields is defined as 
\begin{equation}
T_{\mu\nu}=-\frac{2}{\sqrt{-g}}\frac{\delta S_{M}}{\delta g^{\mu\nu}}~.
\end{equation}
Without specifying the details of the dynamics, we can denote the left hand
sides (lhs) of the Euler-Lagrange equations as 
\begin{equation}
E_{\mu\nu}=\frac{1}{\sqrt{-g}}\frac{\delta S_{G}}{\delta g^{\mu\nu}}
\end{equation}
and 
\begin{equation}
E^{\phi}=\frac{1}{\sqrt{-g}}\frac{\delta S_{G}}{\delta\phi}~,
\end{equation}
the equations of motion being 
\begin{equation}
E_{\mu\nu}=\frac{1}{2}T_{\mu\nu}~,\qquad E^{\phi}=0~.  \label{EOM}
\end{equation}
The lhs' exhibit the following dependencies: 
\begin{align}
E_{\mu\nu} & =E_{\mu\nu}\left(\phi,\partial\phi,\partial^{2}\phi,g,\partial
g,\partial^{2}g\right)~,  \notag \\
E^{\phi} & =E^{\phi}\left(\phi,\partial\phi,\partial^{2}\phi,g,\partial
g,\partial^{2}g\right)~.
\end{align}
Plugging in the continuous fields $g_{\mu\nu}=g_{\mu\nu}^{+}\Theta\left(f%
\right)+g_{\mu\nu}^{-}\Theta\left(-f\right)$ and $\phi=\phi^{+}\Theta\left(f%
\right)+\phi^{-}\Theta\left(-f\right)$, their first derivatives generate
jumps, while the second derivatives terms proportional to $%
\delta\left(f\right)$: 
\begin{align}
E^{\mu\nu} & =\widetilde{E}^{\mu\nu}+\mathcal{E}^{\mu\nu}\alpha\delta\left(f%
\right)~,  \notag \\
E^{\phi} & =\widetilde{E^{\phi}}+\mathcal{E}^{\phi}\alpha\delta\left(f%
\right)~.
\end{align}
Similarly, the energy-momentum tensor allows for a distributional
contribution on $\Sigma$: 
\begin{equation*}
T^{\mu\nu}=\widetilde{T}^{\mu\nu}+\mathcal{T}^{\mu\nu}\alpha\delta\left(f%
\right)~. 
\end{equation*}

The junction conditions are therefore the distributional equations of
motion: 
\begin{equation}
\mathcal{E}^{\mu\nu}=\frac{1}{2}\mathcal{T}^{\mu\nu}~,\qquad\mathcal{E}%
^{\phi}=0~,  \label{junction}
\end{equation}
along with the continuity condition $\left[\phi\right]=\left[g_{\mu\nu}%
\right]=0$.

The scalar equation (\ref{junction}) is simple to be evaluated on $\Sigma$.
The tensor equation (\ref{junction}) can be decomposed with respect to the
oblique basis, employing a $\Sigma$-scalar $\mathcal{E}_{l}$, a $\Sigma$%
-vector $\mathcal{E}_{l}^{a}$ and a $\Sigma$-tensor $\mathcal{E}^{ab}$
defined as 
\begin{equation}
\mathcal{E}^{\mu\nu}=\mathcal{E}_{l}l^{\mu}l^{\nu}+2\mathcal{E}%
_{l}^{a}e_{a}^{(\mu}l^{\nu)}+\mathcal{E}^{ab}e_{a}^{\mu}e_{b}^{\nu}~.
\end{equation}
This decomposition is left unchanged by coordinate transformations on $%
M^{\pm}$. However, only the $\mathcal{E}^{ab}$ part would be nonvanishing,
as the distributional stress-energy tensor $\mathcal{T}^{\mu\nu}$ represents
the intrinsic stress energy of the singular source on the surface. In GR the
vectorial and scalar contributions can be expressed in terms of the
Hamiltonian and diffeomorphism constraints, hence they do not carry new
information. The same has been verified for the simplest scalar-tensor
theories. Therefore, the junction conditions can be rewritten as equations
on $\Sigma$ as 
\begin{equation}
\mathcal{E}^{ab}=\frac{1}{2}\mathcal{T}^{ab}~,\qquad\mathcal{E}^{\phi}=0~.
\label{junctionSigma}
\end{equation}

\section{Brans-Dicke theory}

Perhaps the most well-known scalar-tensor theory is the Brans-Dicke theory,
born from the simple assumption of replacing the gravitational constant with
a scalar field $\phi$. Its Lagrangian in the Jordan frame 
\begin{equation}
L_{BD}=-\frac{\omega}{16\pi\phi}\partial^{\mu}\phi\partial_{\mu}\phi+\frac{%
R\phi}{16\pi}~,
\end{equation}
contains a coupling constant $\omega$. The field equations obtained from the
metric and scalar field variations are 
\begin{align}
8\pi T_{\mu\nu} & =\frac{\omega}{\phi}\left[-\nabla_{\mu}\phi\nabla_{\nu}%
\phi+\frac{1}{2}\left(\nabla\phi\right)^{2}g_{\mu\nu}\right]+\phi G_{\mu\nu}
\notag \\
& -\nabla_{\mu}\nabla_{\nu}\phi+\square\phi g_{\mu\nu}\ ,  \notag \\
0 & =-\frac{\omega}{\phi}\left(\nabla\phi\right)^{2}+\phi
R+2\omega\square\phi  \label{EOMBD}
\end{align}
By exploring the trace of the tensorial equation to eliminate the curvature
scalar, one obtains a Klein-Gordon equation with the trace of the
stress-energy tensor as a source: 
\begin{equation}
\square\phi=\frac{8\pi}{3+2\omega}T~.
\end{equation}

It has been claimed that GR is recovered for the large $\omega$ limit \cite%
{Weinberg}. The GR limit of the Brans-Dicke theory however is intricate,
reducing to GR in the $\omega\rightarrow\infty$ limit only if the trace of
the matter energy-momentum tensor does not vanish. Indeed, in that
particular case (including the vacuum) the asymptotic behaviour in $\omega$
is different \cite{BSen}. This has been explained in \cite{Faraoni} in
terms of the differences in the conformal invariance group modifying $\omega$
in the two cases. In light of this analysis it is not trivial to prove
whether the same limit applies for vacuum, nevertheless it has been assumed
by analysing the Cassini probe data \cite{Cassini}, and stringent constraint $%
\omega>$ $40000$ was set in order the Brans-Dicke theory to survive the
Solar System tests \cite{Freire,CapozzielloFaraoni}.

The Euler-Lagrange expressions for the Brans-Dicke Lagrangian in the Jordan
frame are given by 
\begin{align}
16\pi E_{\mu\nu} & =-\frac{\omega}{\phi}\nabla_{\mu}\phi\nabla_{\nu}\phi+%
\frac{\omega}{2\phi}\left(\nabla\phi\right)^{2}g_{\mu\nu}+\phi
G_{\mu\nu}-\nabla_{\mu}\nabla_{\nu}\phi+\square\phi g_{\mu\nu}~,  \notag \\
16\pi\phi E^{\phi} & =-\frac{\omega}{\phi}\nabla^{\mu}\phi\nabla_{\mu}\phi+%
\phi R+2\omega\square\phi~.
\end{align}
Only the last three terms of $E_{\mu\nu}$ contain second derivatives, so
only those will contribute singular terms.

Due to smoothness in the domains $M^{\pm}$, and continuity through $\Sigma$,
the jump in the derivatives of $g_{\mu\nu}$ and $\phi$ are necessarily
transversal {\cite{BI}}: 
\begin{align}
\left[\partial_{\sigma}g_{\mu\nu}\right] & =\eta n_{\sigma}c_{\mu\nu}~, 
\notag \\
\left[\partial_{\mu}\phi\right] & =\eta n_{\mu}J~.
\end{align}
With these, the singular part of the Einstein-tensor can be expressed {\cite%
{BI} as} 
\begin{equation}
\mathcal{G}_{\mu\nu}=\frac{1}{2}\eta\left(n_{\mu}c_{\nu}+n_{\nu}c_{\mu}-n_{%
\mu}n_{\nu}c-g_{\mu\nu}c^{\dagger}-\epsilon\left(c_{\mu\nu}-cg_{\mu\nu}%
\right)\right)\ ,
\end{equation}
where we introduced the notations 
\begin{equation}
c_{\mu}=c_{\mu\nu}n^{\nu},\ c=c_{\ \mu}^{\mu},\
c^{\dagger}=c_{\mu\nu}n^{\mu}n^{\nu}.
\end{equation}
One may easily check that $\mathcal{G}^{\mu\nu}$ is tangential ($\mathcal{G}%
^{\mu\nu}n_{\nu}=0$) thus it is possible to represent it as an intrinsic $%
\Sigma$-tensor as 
\begin{equation}
\mathcal{G}^{\mu\nu}=\mathcal{G}^{ab}e_{\ a}^{\mu}e_{\ b}^{\nu}\ ,
\end{equation}
this representation being invariant with respect to the choice of
transversal vector $l^{\mu}$.

We find the singular part of the expression $\nabla_{\mu}\nabla_{\nu}\phi-%
\square\phi g_{\mu\nu}$ as 
\begin{equation}
J\eta\left(n_{\mu}n_{\nu}-\epsilon g_{\mu\nu}\right)\ .
\end{equation}
A contraction with $n^{\nu}$ reveals that this term is also tangential.

The tensorial junction equation (\ref{junction}) then reads 
\begin{align}
8\pi\mathcal{T}^{\mu\nu} & =\frac{1}{2}\phi\eta\left(n^{\mu}c^{\nu}+n^{%
\nu}c^{\mu}-n^{\mu}n^{\nu}c-g^{\mu\nu}c^{\dagger}-\epsilon\left(c^{\mu%
\nu}-cg^{\mu\nu}\right)\right)  \notag \\
& -J\eta\left(n^{\mu}n^{\nu}-\epsilon g^{\mu\nu}\right)\ .
\label{eq:junctionext}
\end{align}
In what follows we will express this equation as an intrinsic $\Sigma$%
-tensor equation.

A convenient basis of the tangent spaces of $M$ along $\Sigma$ is $%
\left(l^{\mu},e_{\ a}^{\mu}\right)$, with the dual frame $\left(\eta
n_{\mu},\theta_{\ \mu}^{a}\right)$, where $\theta_{\ \mu}^{a}$ obeys the
relations 
\begin{equation}
\theta_{\ \mu}^{a}e_{\ b}^{\mu}=\delta_{b}^{a}~,\qquad\theta_{\
\mu}^{a}l^{\mu}=0\ .
\end{equation}
Unlike $e_{\ a}^{\mu}$, the covector fields $\theta_{\ \mu}^{a}$ depend on
the choice of $l^{\mu}$. For any vector $X^{\mu}$ along $\Sigma$, the
contraction $\theta_{\ \mu}^{a}X^{\mu}$ provides the tangential components
of $X$. Further, when $X$ is purely tangential, this contraction simply
provides its $l^{\mu}$-independent components in the coordinate frame
adapted to $\Sigma$. This will be explored in identifying $\mathcal{E}^{ab}$
from $\mathcal{E}^{\mu\nu}$, when $\mathcal{E}^{a}=0=\mathcal{E}$, and
rewriting Eq. (\ref{eq:junctionext}) accordingly. In doing so we will
explore the jump of the extrinsic curvature.

The extrinsic curvature of a space-like or time-like surface with normal $%
n^{\mu}$ is defined as $K_{ab}=\frac{1}{2}e_{\ a}^{\mu}e_{\ b}^{\nu}\mathcal{%
L}_{n}g_{\mu\nu}$. For null hypersurfaces this quantity does not carry
transverse information (as $n^{\mu}$ becomes tangential), hence we replace
it with the analogous transverse curvature {\cite{BI}} 
\begin{equation}
\mathcal{K}_{ab}=\frac{1}{2}e_{\ a}^{\mu}e_{\ b}^{\nu}\mathcal{L}%
_{l}g_{\mu\nu}\ .
\end{equation}
The jump of the transverse curvature is 
\begin{equation}
\left[\mathcal{K}_{ab}\right]=\frac{1}{2}e_{\ a}^{\mu}e_{\
b}^{\nu}c_{\mu\nu}\equiv\frac{1}{2}c_{ab}\ .
\end{equation}
To show how this relates to the full $c_{\mu\nu}$, we decompose the latter
with respect to the dual frame 
\begin{equation}
c_{\mu\nu}=c^{n}n_{\mu}n_{\nu}+2c_{a}^{n}n_{(\mu}\theta_{\ \nu)}^{a}+2\left[%
\mathcal{K}_{ab}\right]\theta_{\ \mu}^{a}\theta_{\ \nu}^{b}\ .
\label{eq:cdecomp}
\end{equation}
We also decompose the inverse metric with respect to the vector frame 
\begin{equation}
g^{\mu\nu}=\epsilon\eta^{2}l^{\mu}l^{\nu}+2\eta n^{a}l^{(\mu}e_{\
a}^{\nu)}+h_{\ast}^{ab}e_{\ a}^{\mu}e_{\ b}^{\nu}\ ,
\end{equation}
where $n^{a}=n^{\mu}\theta_{\mu}^{a}$ is the ``tangential\textquotedblright
\ part of $n$ in the $n^{\mu}=n_{l}l^{\mu}+n^{a}e_{\ a}^{\mu}$
decomposition, and $h_{\ast}^{ab}$ is a pseudo-inverse metric on $\Sigma$,
defined as $h_{\ast}^{ab}=g^{\mu\nu}\theta_{\mu}^{a}\theta_{\nu}^{b}$ (it
becomes the inverse for non-null hypersurfaces if the transverse is chosen
as the normal).

Because Eq. (\ref{eq:junctionext}) is tangential (as can be seen by
contracting with $n_{\mu}$), contracting with $\theta_{\ \mu}^{a}\theta_{\
\nu}^{b}$ gives the intrinsic components of the tensorial junction condition 
\begin{align}
\mathcal{T}^{ab} & =\frac{\phi\eta}{8\pi}\left[\mathcal{K}_{cd}\right]%
\left(h_{\ast}^{ac}n^{b}n^{d}+h_{\ast}^{bc}n^{a}n^{d}-h_{%
\ast}^{cd}n^{a}n^{b}-h_{\ast}^{ab}n^{c}n^{d}\right.  \notag \\
&
\left.-\epsilon\left(h_{\ast}^{ac}h_{\ast}^{bd}-h_{\ast}^{ab}h_{\ast}^{cd}%
\right)\right)-\frac{J\eta}{8\pi}\left(n^{a}n^{b}-\epsilon
h_{\ast}^{ab}\right)  \label{eq:junct_tensor_bd}
\end{align}
This is the analogue of the Lanczos equation of GR and it is valid for
arbitrary junction surface $\Sigma$.

The non-null GR limit is readily obtained with $\phi=G^{-1}$ and $J=0$, and
further simplified by the choice $l^{\mu}=n^{\mu}$. This gives $n^{a}=0$, $%
\eta=\epsilon$, $\mathcal{K}_{ab}=K_{ab}$ and $h_{\ast}^{ab}=h^{ab}$.
Inserting these into Eq. (\ref{eq:junct_tensor_bd}) gives 
\begin{equation}
\mathcal{T}^{ab}=-\frac{1}{8\pi G}\left(\left[K^{ab}\right]-\left[K\right]%
h^{ab}\right)~,
\end{equation}
the familiar Lanczos equation of GR.

Next we consider the scalar junction condition (\ref{junction}). In the
scalar equation of motion (\ref{EOMBD}) only the terms $\phi
R+2\omega\square\phi$ contain second derivatives, only they contribute the
singular parts 
\begin{equation}
\mathcal{E}_{\phi}=\frac{1}{16\pi}\left(\phi\eta\left(c^{\dagger}-\epsilon
c\right)+2\omega\eta\epsilon J\right)~.
\end{equation}
As a scalar equation, this is already intrinsic to $\Sigma$. Proceeding as
in the case of the tensorial equation, we explore Eq. (\ref{eq:cdecomp}) to
express $c^{\dagger}$ and $c$ in terms of $c^{n}$, $c_{a}^{n}$ and $c_{ab}=2%
\left[\mathcal{K}_{ab}\right]$. After simplification and inserting into $%
\mathcal{E}_{\phi}=0$ we get 
\begin{equation}
0=\phi\eta\left[\mathcal{K}_{ab}\right]\left(n^{a}n^{b}-\epsilon
h_{\ast}^{ab}\right)+2\omega\eta\epsilon J~.  \label{eq:junct_scalar_bd}
\end{equation}
This, together with Eq. (\ref{eq:junct_tensor_bd}) constitute the junction
equations in Brans-Dicke theory.

The non-null limit (with the choice $l^{\mu}=n^{\mu}$ and consequences as
described earlier) of these junction conditions arises as 
\begin{eqnarray}
\mathcal{T}^{ab} & = & -\frac{\phi}{8\pi}\left(\left[K^{ab}\right]-\left[K%
\right]h^{ab}-Jh^{ab}\right)~, \\
0 & = & -\frac{1}{8\pi}\left(\phi\left[K\right]-2\omega J\right)~.
\end{eqnarray}
These junction conditions derived in the Jordan frame correspond to the one
scalar case of those obtained in Einstein frame in the context of
multi-scalar tensor theories of gravity in \cite{BarrabesBressange}.

\section{The null case}

The junction equations derived so far are applicable to any hypersurface,
irrespective of its causal character. In what follows, we assume $\Sigma$ a
null hypersurface, so the normal vector becomes also tangential. Hence the
choice $l^{\mu}=n^{\mu}$ leads to a degenerate situation, and another
simplifying assumption for the transverse vector should be made. This is
covered by our forthcoming analysis, which in turn closely follows the one
presented for GR by Poisson \cite{Poisson}. We will chose an autoparallel
normal vector, denoted $N^{\mu}$ (satisfying $N^{\nu}\nabla_{\nu}N^{\mu}=%
\kappa N^{\mu}$) and we will use a coordinate system adapted to $N^{\mu}$.
The parameter along the integral curves of $N^{\mu}$ is $\lambda$, one of
the coordinates on $\Sigma$. We denote the two additional coordinates $%
\left\{ \sigma^{A}\right\} $ (capital latin indices taking the values $2,3$%
), labelling the geodesic integral curves of $N^{\mu}$. The transverse
vector, denoted $L^{\mu}$ is chosen \cite{Poisson,HE} as 
\begin{equation}
L_{\mu}L^{\mu}=0,\ L^{\mu}N_{\mu}=1,\ L_{\mu}e_{\ A}^{\mu}=0~.
\label{pseudoorthonormalbasis}
\end{equation}
Here $e_{\ A}^{\mu}=\partial x^{\mu}/\partial\sigma^{A}$ are the coordinate
basis fields associated to $\sigma^{A}$. If $e_{~A}^{\mu}$ also obey
orthonormality relations, the basis (\ref{pseudoorthonormalbasis}) is
pseudoorthonormal. The induced metric in this basis is manifestly
two-dimensional: 
\begin{equation}
h_{11}=g_{\mu\nu}N^{\mu}N^{\nu}=0,\ h_{1A}=g_{\mu\nu}N^{\mu}e_{~A}^{\nu}=0,\
q_{AB}\equiv h_{AB}=g_{\mu\nu}e_{\ A}^{\mu}e_{\ B}^{\nu}\ .
\end{equation}
It is easy to check that $q_{AB}$ (as opposed to $h_{ab}$) is non-degenerate,
and a unique inverse $q^{AB}$ exists. If we define the dual frame $%
e_{\mu}^{\ A}=g_{\mu\nu}q^{AB}e_{\ B}^{\nu}$, it obeys 
\begin{equation}
e_{\ \mu}^{A}e_{\ B}^{\mu}=q_{BC}q^{AC}=\delta_{B}^{A}~.
\end{equation}
The dual frame of $\left(L^{\mu},N^{\mu},e_{\ A}^{\mu}\right)$ is $%
\left(N_{\mu},L_{\mu},e_{\mu}^{\ A}\right)$.

With it the components of the normal (defined as $n^{a}=n^{\mu}\theta_{%
\mu}^{a}$) become $N^{\lambda}=N^{\mu}L_{\mu}=1$ and $N^{A}=N^{\mu}e_{\mu}^{%
\ A}=0$, hence $n^{a}=\delta_{\lambda}^{a}$. The normal $n^{a}\equiv e_{\
\lambda}^{a}=\delta_{\lambda}^{a}$ is therefore a first basis vector on the
tangent space of $\Sigma$, the other two being $e_{\ A}^{a}=\delta_{~A}^{a}$%
. The pseudo-inverse metric has the components 
\begin{equation*}
h_{\ast}^{\lambda\lambda}=g^{\mu\nu}L_{\mu}L_{\nu}=0,~h_{\ast}^{\lambda
A}=g^{\mu\nu}L_{\mu}e_{\nu}^{\ A}=0,~h_{\ast}^{AB}=g^{\mu\nu}e_{\mu}^{\
A}e_{\nu}^{\ B}=q^{AB}~. 
\end{equation*}
By inserting these into Eq. (\ref{eq:junct_tensor_bd}) and substituting $%
\eta=1$ and $\epsilon=0$, we obtain the tensorial junction condition 
\begin{equation}
\mathcal{T}^{ab}=\rho e_{\ \lambda}^{a}e_{\ \lambda}^{b}+j^{A}\left(e_{\
A}^{a}e_{\ \lambda}^{b}+e_{\ \lambda}^{a}e_{\ A}^{b}\right)+pq^{AB}e_{\
A}^{a}e_{\ B}^{b}~,
\end{equation}
where 
\begin{equation}
\rho=-\frac{\phi}{8\pi}\left[\mathcal{K}_{AB}\right]q^{AB}-\frac{J}{8\pi}
\end{equation}
is the surface density of the layer, 
\begin{equation}
j^{A}=\frac{\phi}{8\pi}\left[\mathcal{K}_{B\lambda}\right]q^{AB}
\end{equation}
the surface current of the layer, and 
\begin{equation}
p=-\frac{\phi}{8\pi}\left[\mathcal{K}_{\lambda\lambda}\right]
\end{equation}
the isotropic pressure of the layer. With $\phi=G^{-1}$ and $J=0$ the
corresponding result derived in \cite{Poisson} is reobtained.

Finally the scalar junction equation (\ref{eq:junct_scalar_bd}) simplifies
to $0=\phi\left[\mathcal{K}_{\lambda\lambda}\right]$, implying that 
\begin{equation*}
p=0~, 
\end{equation*}
thus the presence of a continuous scalar field contributing to the
expressions of the surface density, current and pressure does not allow for
an isotropic surface pressure on the null layer. This result is consistent with the pressurelessness condition derived in \cite{BarrabesBressange} for a multiscalar generalization of Brans-Dicke theory in the Einstein frame.

\section{Summary}

The junction of space-time regions with different geometrical
characteristics is an important task in all geometric theories of gravity.
It is of particular interest when the separating hypersurface is singular,
carrying distributional energy-momentum tensor. The junction formalism is
complicated if the hypersurface is null. We have derived the generic
junction conditions for Brans-Dicke theory in the Jordan frame, as this is
the frame in which the matter energy-momentum conservation holds (the
physical frame). We explored a formalism based on a transverse vector,
rather than normal, which can be applied to any type of hypersurface. Then
we considered the particular cases of (i) non-null hypersurfaces, obtaining
the generalisations of the Lanczos equation, in which the jump of the
extrinsic curvature is sourced by both the distributional energy-momentum
tensor and by the jump in the transverse derivative of the scalar, and ii)
null hypersurfaces, which represent shock-waves propagating with the speed
of light. In the latter case the distributional source is decomposed into
surface density, current and pressure. The latter however ought to vanish by
virtue of the scalar junction condition. A similar result derived
as a traceless requirement for the distributional energy-momentum source in
the Einstein frame is a remarkable example of
the frame independence of a physical result.

Confronting these results with previous ones derived for non-null
hypersurfaces in the Einstein frame, as well as their generalisation for the
so-called Generalised Brans-Dicke theory (given by the Lagrangian $L_{GBD}=%
\frac{1}{2}F\left( \phi \right) R+B\left( \phi \right) X-2G\left( \phi
\right) \square \phi X$, where $F,B,G$ are arbitrary smooth functions of $%
\phi $ and $X$ are in progress and will be presented elsewhere.%

\acknowledgments{L.\'A.G. thanks Shinji Mukohyama for inspiring discussions on the topic.
This work was supported by the Hungarian National Research Development
and Innovation Office (NKFI) in the form of the grant 123996. 
The authors thank the organisers of the Bolyai-Gauss-Lobachevsky Conference for partial support of their participation.}

\authorcontributions{All authors contributed equally to this work.}

\conflictsofinterest{The authors declare no conflict of interest.} 



\reftitle{References}


\begin{thebibliography}{999}

\bibitem{GW150915} LIGO Scientific Collaboration, Virgo Collaboration,
Observation of Gravitational Waves from a Binary Black Hole Merger. \emph{%
Phys. Rev. Lett.} \textbf{2016}, \emph{116}, 061102

\bibitem{GW151226} LIGO Scientific Collaboration, Virgo Collaboration,
GW151226: Observation of Gravitational Waves from a 22-Solar-Mass Binary
Black Hole Coalescence. \emph{Phys. Rev. Lett.} \textbf{2016,} \emph{116},
241103

\bibitem{GW170104} LIGO Scientific Collaboration, Virgo Collaboration,
GW170104: Observation of a 50-Solar-Mass Binary Black Hole Coalescence at
Redshift 0.2. \emph{Phys. Rev. Lett.} \textbf{2017}, \emph{118}, 221101

\bibitem{GW170608} LIGO Scientific Collaboration, Virgo Collaboration,
GW170608: Observation of a 19-Solar-Mass Binary Black Hole Coalescence.
arXiv:1711.05578 {[}astro-ph.HE{]} \textbf{2017}

\bibitem{GW170814} LIGO Scientific Collaboration, Virgo Collaboration,
GW170814: A Three-Detector Observation of Gravitational Waves from a Binary
Black Hole Coalescence. \emph{Phys. Rev. Lett.} \textbf{2017}, \emph{119},
141101

\bibitem{GW170817} LIGO Scientific Collaboration, Virgo Collaboration,
GW170817: Observation of Gravitational Waves from a Binary Neutron Star
Inspiral. \emph{Phys. Rev. Lett.} \textbf{2017,} \emph{119}, 161101

\bibitem{GRtests} LIGO Scientific Collaboration, Virgo Collaboration, Tests
of General Relativity with GW150914. \emph{Phys. Rev. Lett.} \textbf{2016,} 
\emph{116}, 221101

\bibitem{GWGRB} LIGO Scientific Collaboration, Virgo Collaboration, Fermi
Gamma-ray Burst Monitor, and INTEGRAL, Gravitational Waves and Gamma-Rays
from a Binary Neutron Star Merger: GW170817 and GRB170817A. \emph{The
Astrophysical Journal Letters} \textbf{2017,} \emph{848}, L13

\bibitem{Horndeski} Horndeski, G. W., Second-Order Scalar-Tensor Field
Equations in a Four-Dimensional Space.\emph{\ Int. J. Theor. Phys.} \textbf{%
1974}, \emph{10} , 363

\bibitem{DGSZ} Deffayet, C., Gao, X., Steer D.A., Zahariade, G., From
k-essence to generalized Galileons. \emph{Phys. Rev. D} \textbf{2011}, \emph{%
84}, 064039

\bibitem{CapozzielloFaraoni} Capozziello, S., Faraoni, V. The variational
principle and the field equations of Brans-Dicke gravity. In \emph{Beyond
Einstein Gravity: A Survey of Gravitational Theories for Cosmology and
Astrophysics}; Springer, 2011; pp. 59-61, ISBN
978-94-007-0164-9

\bibitem{Dicke} Dicke, R. H., Mach's principle and invariance under
transformation of units. \emph{Phys. Rev.} \textbf{1962}, \emph{125}, 2163

\bibitem{FaraoniNadeau} Faraoni, V.,Nadeau, S., The (pseudo)issue of the
conformal frame revisited. \emph{Phys. Rev. D. }\textbf{2007}, \emph{75},
023501

\bibitem{FaraoniGunzig} Faraoni, V., Gunzig, E., Einstein frame or Jordan
frame? \emph{Int. J. Theor. Phys.} \textbf{1999}, \emph{38}, 217

\bibitem{MagnanoSokolowki} Magnano, G., Sokolowski, L.M., On Physical
Equivalence between Nonlinear Gravity Theories and a General\textendash
Relativistic Self\textendash Gravitating Scalar Field. \emph{Phys. Rev. D.} 
\textbf{1994}, \emph{50}, 5039

\bibitem{BarrabesBressange} Barrab\`{e}s, C, Bressange, G. F., Singular
hypersurfaces in scalar - tensor theories of gravity. \emph{Class. Quantum
Grav.} \textbf{1997}, \emph{14}, 805

\bibitem{Israel} Israel, W, Singular hypersurfaces and thin shells in
general relativity. \emph{Nuovo Cim. B} \textbf{1966}, \emph{44}, 1

\bibitem{BI} Barrab\`{e}s, C, Israel, W, Thin shells in general relativity and
cosmology: The lightlike limit. \emph{Phys. Rev. D} \textbf{1991}, \emph{43}%
, 1129

\bibitem{Mukohyama} Mukohyama, S., Doubly covariant action principle of
singular hypersurfaces in general relativity and scalar-tensor theories. 
\emph{Phys. Rev. D }\textbf{2001}, $\emph{65}$, 024028

\bibitem{SchmidtWang} Schmidt, H., Wang, A., Plane domain walls when coupled with the Brans-Dicke scalar field.
\emph{Phys. Rev. D} \textbf{1993}, \emph{47}, 4425

\bibitem{LetelierWang} Letelier, P. S., Wang, A., Spherically symmetric thin shells in Brans-Dicke theory of gravity. \emph{Phys. Rev. D} \textbf{1993}, \emph{48}, 631

\bibitem{SakaiMaeda1} Sakai, N., Maeda, K., Bubble dynamics and space-time structure in extended inflation. \emph{Phys. Rev. D} \textbf{1993}, \emph{48}, 5570

\bibitem{SakaiMaeda2} Sakai, N., Maeda, K., Bubble dynamics in generalized Einstein theories. \emph{Prog. Theor. Phys.} \textbf{1993}, \emph{90}, 1001

\bibitem{PS} Padilla, A., Sivanesan, V., Boundary terms and junction
conditions for generalized scalar-tensor theories. \emph{J. High Energ. Phys.%
} \textbf{2012}, \emph{JHEP08}, 122

\bibitem{NishiKobayashi} Nishi, S.,Kobayashi, T., Tanahashi, N., Yamaguchi,
M.,Cosmological matching conditions and galilean genesis in Horndeski's
theory. \emph{J. Cosmol. Astropartic. Phys.} \textbf{2014}, \emph{JCAP03},
008

\bibitem{GerochTraschen} Geroch, R., Traschen, J.H., Strings and Other
Distributional Sources in General Relativity. \emph{Phys. Rev. D} \textbf{%
1987}, \emph{36}, 1017

\bibitem{ClarkeDray} Clarke, C.J.S., Dray, T., Junction conditions for null
hypersurfaces. \emph{Class. Quantum Grav.} \textbf{1987}, \emph{4}, 265

\bibitem{Ostrogradsky} Ostrogradsky, \emph{M\`{e}m. de l'acad. de St. P\`{e}tersbourg%
} \textbf{1850}, \emph{4}, 385

\bibitem{Woodard} Woodard, R.P., The Theorem of Ostrogradsky.
arXiv:1506.02210 {[}hep-th{]} \textbf{2015}

\bibitem{KobayashiYamaguchi} Kobayashi, T., Yamaguchi, M., Yokoyama, J., Generalized G-inflation: Inflation with the most general second-order field equations. \emph{Prog. Theor. Phys.} \textbf{2011}, \emph{126}, 511

\bibitem{DeFeliceTsujikawa} De Felice, A., Tsujikawa, S., Conditions for the cosmological viability of the most general scalar-tensor theories and their applications to extended Galileon dark energy models. \emph{JCAP} \textbf{2012}, \emph{1202}, 007

\bibitem{Many} Baker. T., Bellini, E., Ferreira P.G., Lagos M., Noller J., Sawicki I., Strong constraints on cosmological gravity from GW170817 and GRB 170817A. \emph{Phys. Rev. Lett.} \textbf{2017} \emph{119} 251301

\bibitem{EZ} Ezquiaga, J.M., Zumalacarregu, M., Dark Energy after GW170817: dead ends and the road ahead. \emph{Phys. Rev. Lett} \textbf{2017}, \emph{119} 251304

\bibitem{CreminelliVernizzi} Creminelli, P., Vernizzi, F., Dark Energy after GW170817 and GRB170817A. \emph{Phys. Rev. Lett.} \textbf{2017}, \emph{119}, 251302

\bibitem{Weinberg} Weinberg, S. The Brans-Dicke theory. In \emph{Gravitation
and Cosmology}; John Wiley \& Sons inc.; 1972;
pp. 157-160, ISBN 0-471-92567-5

\bibitem{BSen} Banerjee N., Sen, S., Does Brans-Dicke theory always yield
general relativity in the infinite $\omega$ limit? \emph{Phys. Rev. D} 
\textbf{1997},\emph{\ 56}, 1334

\bibitem{Faraoni} Faraoni, V., The $\omega\rightarrow\infty$ limit of
Brans-Dicke theory. \emph{Phys. Lett. A} \textbf{1998}, $\emph{245}$, 26

\bibitem{Cassini} Bertotti, B., Iess, L., Tortora, P., A test of general
relativity using radio links with the Cassini spacecraft. \emph{Nature} 
\textbf{2003}, \emph{425}, 374

\bibitem{Freire} Freire, P.C.C., Wex, N., Esposito-Farèse, G., Verbiest, J. P. W.,
Bailes, M., Jacoby, B.A., Kramer, M., Stairs, I.H.,
Antoniadis, J., Janssen, G.H., The relativistic pulsar–white dwarf binary PSR J1738+0333 – II. The
most stringent test of scalar–tensor gravity. \emph{Mon. Not. R. Astron. Soc.} \textbf{2012}, \emph{423}, 3328

\bibitem{Poisson} Possion, E. Null Shells. In \emph{A Relativist's Toolkit:
The Mathematics of Black-Hole Mechanics}; Cambridge University
Press, 2004; pp. 98-104, ISBN 0-521-83091-5

\bibitem{HE} Hawking, S. W., Ellis, G. F. R., Null curves. In \emph{The
Large-Scale Structure of Spacetime}; Cambridge University
Press, 1973; pp. 86-87, ISBN 0-521-09906-4
\end{thebibliography}
\end{document}